# Spiral morphology in an intensely star-forming disk galaxy more than 12 billion years ago

Takafumi Tsukui[1,2]*, Satoru Iguchi[1,2]

**Affiliations:**

[1]Department of Astronomical Science, SOKENDAI (The Graduate University for Advanced Studies), 2-21-1 Osawa, Mitaka, Tokyo, Japan.

[2]National Astronomical Observatory of Japan, National Institute of Natural Sciences, 2-21-1 Osawa, Mitaka, Tokyo, Japan.

*Corresponding author. Email: t.tsukui@grad.nao.ac.jp

**Abstract:**

Spiral galaxies have distinct internal structures including a stellar bulge, disk and spiral arms. It is unknown when in cosmic history these structures formed. We analyze observations of BRI 1335-0417, an intensely star-forming galaxy in the distant Universe, at redshift 4.41. The [C II] gas kinematics show a steep velocity rise near the galaxy center and have a two-armed spiral morphology that extends from about 2 to 5 kiloparsecs in radius. We interpret these features as due to a central compact structure, such as a bulge, a rotating gas disk and either spiral arms or tidal tails. These features had been formed within 1.4 billion years after the Big Bang, long before the peak of cosmic star formation.

Spiral galaxies contain distinct internal structures: a stellar bulge; a flat, extended rotating disk; and spiral arms that extend through the disk. Observations have established that the total star formation rate in the Universe rises steeply after the Big Bang, reaches a peak at redshift $z$ between 1.5 and 3, and declines steadily since then (*1*). Optical observations provide evidence that stellar bulges are already formed in some massive galaxies by $z$ of 0.5 to 2.5 (*2*). Spiral structure has been reported in galaxies at redshifts of $z$=2.18 (*3*), $z$=2.01 (*4*) and $z$=2.54 (*5*). Observations with the Atacama Large Millimeter/submillimeter Array (ALMA) have shown spiral structure in two galaxies at $z$=2.32 and $z$=2.67 (*6*), dynamically cold disks in galaxies at redshift $z \sim 4$ (*7–10*) and compact central structures, like a bulge, at $z$ of 4 to 5 (*7, 10*). It is currently unclear whether the later appearance of spiral structure is due to different formation times or observational limitations.

BRI 1335-0417 (QSO J1338-0432), a galaxy at redshift $z = 4.4074 \pm 0.0015$ (*11*), 1.4 Gyr after the Big Bang. It is classified as a Hyper Luminous InfraRed Galaxy (HyLIRG) with a luminosity of $3.1 \times 10^{13} L_\odot$ ($L_\odot$ is the luminosity of the Sun) at far-infrared wavelengths (*12*) and a high star formation rate of $5 \pm 1 \times 10^3 M_\odot \text{yr}^{-1}$ ($M_\odot$ is the mass of the Sun), estimated from modeling the spectral energy distribution (*13*).

We analyze archival ALMA observations of BRI 1335-0417 in the gas emission line [C II] (rest frame frequency of 1900.5369 GHz, redshifted to around 351 GHz in the observed frame). The data were produced using standard procedures (*14*). [C II] line emission from this object was detected in previous observations at lower spatial resolution, insufficient to determine the gas distribution and kinematics (*15*). Figure 1A-B shows the ALMA [C II] intensity and velocity images, with a spatial resolution of of 0.20" × 0.16" full width half maximum (FWHM), which is equivalent to ~ 1.33 × 1.11 kiloparsecs (kpc) at this redshift. Figure 1C shows the dust continuum image at a central frequency of 1863 GHz in rest frame with spatial resolution of 0.17" × 0.14" (~ 1.16 × 0.95 kpc). The root mean square noise in each velocity channel (width of 20 km s$^{-1}$) is about 0.35 millijansky per beam (mJy beam$^{-1}$), while that of the dust continuum map is about 0.036 mJy beam$^{-1}$. Figure 1D shows a composite image of the gas velocities, produced from [C II] velocity channel maps (Fig. 2) that show the [C II] line intensity images at each velocity.

The [C II] line and dust continuum images show a central compact structure surrounded by more extended emission which could indicate a disk. A spiral morphology is apparent in Fig. 1D, with northern and southern arms extending from 2 to 5 kpc in radius. This spiral morphology is more apparent in the [C II] velocity channel maps (Fig. 2). We compare these maps to previously-published CO ($J$=2→1) line images (where $J$ is the rotational quantum number) which showed different spatial distribution depending on the antenna configuration used for the observations. One CO image showed a structure with a bright central component and a weaker northern component (*16*), while the another showed a complex structure with multiple components (*17*). These multiple CO components correspond (in position and velocity) to the northern spiral feature, southern spiral feature and central disk seen in the [C II] images. [C II] line emission arises from multiple phases of the interstellar medium, including ionized gas, atomic gas and molecular gas, while CO ($J$=2→1) line emission traces only dense molecular gas (*18*). The difference in [CII] and CO distributions is also caused by the difference in sensitivity and spatial frequency coverage between their observations. In addition, in [C II] line emission, we could not find any evidence of bipolar outflow which is reported in CO ($J$=7→6) line (*14*).

The [C II] line velocity rises steeply in the center of the galaxy, reaching about 240-260 km s$^{-1}$, indicating the presence of a centrally concentrated mass distribution. The velocity is approximately constant at 70-110 km s$^{-1}$ in the outer parts of the galaxy along the kinematic major axis, as seen in a position-velocity diagram (PVD; Fig. 1E). A steeply rising and/or constant velocity curve has been measured in other high redshift galaxies (*7*, *8*, *10*). Such a velocity curve has been observed in nearby spiral galaxies, produced by the gravitational potential of the internal structures (*19*). The [C II] PVD of BRI 1335-0417 indicates the presence of a central compact structure, such as a stellar bulge, and an extended disk. The rise in velocity appears to be steeper on the northern side than the southern side. The decrease in velocity until a radius of 0.3" is seen in the southern side, while the slight increase with radius is seen in the northern side. These features could be due to streaming motions along spiral arms, warping of the disk, recent gas accretion (*10*) and/or a lopsided central gas distribution. These disturbed features are common in nearby galaxies, and have also been found in high-redshift galaxies at $z \sim 4$ (*8*, *10*). Figure 1F shows the velocity dispersions extracted at each position along the kinematic major axis. The velocity dispersions in the outer parts are consistent with a constant $\sim 70$ km s$^{-1}$, but increase up to $\sim 140$ km s$^{-1}$ in the center. Such central increases are known to be affected by the beam smearing effect (*20*): because the spatial resolution is not sufficient to resolve the intrinsic velocity gradient, it is averaged over the beam width, so the observed velocity dispersion increases. We estimate an intrinsic velocity dispersion $\sigma_v$ of $71^{+14}_{-11}$ km s$^{-1}$ (Fig. 1F) using a method (*21*) that extracts the velocity dispersions at the outer region of the disk, where the beam smearing is less severe.

We used Fourier analysis (*3*) to decompose the [C II] intensity image (Fig. 1A) into logarithmic spirals with *m* arms (Fig. 3A) (*14*). The best-fitting model is a two-armed spiral morphology with a pitch angle of $26.7^{+4.1}_{-1.6}°$ (95% confidence level). Figure 3B shows this model overlain on the [C II] line intensity image after correction for the inclination angle of 37.8° (i.e. de-projected to be viewed as face-on). The disk inclination was estimated from the axis ratio of the dust continuum image (Table S1) (*14*). The amplitude of the *m*=2 signal is 6.0 times the noise, while the other modes *m*=1, 3, 4 are have signal-to-noise ratios of 4.0, 2.5 and 2.4 respectively (Fig. 3A). The *m* = 1 mode corresponds to a one-arm spiral or lopsided morphology. The broad signal in *m* = 1 mode (Fig. 3A) is due to the difference in length of two arms (Fig. 3B). The *m* = 3 and 4 modes correspond to a triangular and boxy shape respectively, which could be produced by a stellar bar structure.

We estimate the inclination-corrected rotational velocity $v_{rot}$ is $179^{+25}_{-18}$ km s$^{-1}$ at radius of 0.3", calculated using the flat part of the PVD (Fig. 1E). The ratio of $v_{rot}$ to the intrinsic velocity dispersion $\sigma_v$ quantifies the rotational support of the disk (*20*). Our estimated ratio $v_{rot}/\sigma_v$ of $2.5^{+0.6}_{-0.4}$ indicates that the [C II] gas disk of this galaxy is rotation-dominated. Similar values have been found for other galaxies observed at $z \sim 4$ (*8*, *9*). Higher ratios have also been reported at redshift of 4 to 5 (*7*, *10*). Additional evidence of rotating disk is symmetric velocity field across the kinematic minor axis, seen in this object (Fig. 1B). Ordered rotation has been found in kinematic data of several galaxies at redshift ~ 4 (*22*), with such symmetric velocity fields.

We estimate the mass distribution by modeling gas motions under a gravitational potential composed of a central compact structure and an extended, rotating disk (*14*). The best-fitting effective radius of the compact structure $R_e$ is less than 1.3kpc (at 95% confidence interval) and its mass is $5.2 \times 10^9$ to $3.0 \times 10^{10}$ $M_\odot$ depending on the value of $R_e$. These results are compatible with the typical effective radius and mass of stellar bulges in nearby galaxies (*23*). The modeled mass of the disk $4.9^{+1.7}_{-2.5} \times 10^{10}$ $M_\odot$ is consistent with the molecular gas mass of $5.1 \times 10^{10}$ $M_\odot$ (*16*) estimated from the CO ($J=2\rightarrow1$) luminosity (assuming that BRI 1335-0417 has metal abundance of the sun). We computed the Toomre parameter $Q$ of the gas disk, which describes the gravitational stability of the disk against perturbations (*14*). We find $Q$ is less than 1 throughout the outer part of the disk (Fig. S6), indicating the gas disk is susceptible to gravitational collapse, star formation (*7–9*) and spontaneous formation of spiral structure (*3, 24*).

Studies of nearby galaxies have shown that bulges and supermassive black holes coexist in massive galaxies, and that there is a strong correlation between their properties (*25*). The mass of the central black hole in BRI 1335-0417 has previously been estimated as $M_{BH} \sim 6 \times 10^9$ $M_\odot$ (*26*), although there are systematic effects that could bias this value (*27*). Our modeled compact structure mass $5.2 \times 10^9$-$3.0 \times 10^{10}$ $M_\odot$ is consistent with the estimated $M_{BH}$. The ratio black hole mass to the compact structure mass is 1-0.2, which is higher than the black hole to bulge mass ratio of 0.001 to 0.002 observed in nearby galaxies (*25*) and includes the possibility of a supermassive black hole without a bulge. There is tentative evidence that the black hole-bulge mass ratio in high redshift galaxies could be higher than at low redshift, but the data are inconclusive (*28*).

Dusty starburst galaxies like BRI 1335-0417 are the progenitors of present-day massive elliptical galaxies, which are dominated by old stellar populations (*7*). This transformation requires contraction of the disk by angular momentum transport through gravitational torques driven by non-axisymmetric structures (*29*). The spiral morphology in BRI 1335-0417 is present at $z = 4.41$, long before the peak of cosmic star formation. Spiral structure could influence the evolution of this galaxy: the angular momentum could be redistributed, triggering gas inflow into the center of the galaxy and driving intense star formation there (*29*). Such gas inflow could not be directly detected in our data due to the limited signal to noise ratio.

The two-armed spiral morphology observed in BRI 1335-0417 can be induced in the disk through tidal interactions (*3*, *24*). Because this galaxy has a rotating disk and a compact structure at its center, sufficient time for their formation is likely to have passed after the major merger event. Cosmological simulations at redshift of 6 (*30*) show that spiral structure on a rotating disk can appear when the disk is relaxed after a merger event. The two-armed spiral morphology in BRI 1335-0417 extends out to 5kpc (Fig. 1A), is connected to the galaxy center (Fig. 1D) and associated with the rotating disk spatially and in velocity (Fig. 2 and Fig. 1E). Tidal-tail features in a galaxy merger are expected occur on larger scales (>10kpc) (*6*, *31*). We therefore disfavor, but cannot completely rule out, a tidal-tail origin for the observed spiral morphology. Alternatively, a stellar bar structure may play a role to form two-armed spiral structure (*24*, *32*) in this object. Figures 1A and 1C show non-axisymmetric structures, which could be due to a stellar bar, and the spiral arms appear to begin from the edge of this bar structure, as seen in nearby disk galaxies and numerical simulation (*32*).

The high star formation rates of *z*>4 galaxies like BRI 1335-0417 are commonly explained as the result of major mergers, which could produce distorted galactic kinematics. We find that BRI 1335-0417 has only slightly disturbed, rotation dominated kinematics, which can be well described by a rotating disk model (*14*). This suggests that the high star formation rate must have been maintained long enough for the disk to form after any major merger event. The *Q* parameter shows the outer disk of BRI 1335-0417 is unstable, which could be caused by gas accretion along large-scale filaments of the cosmic web (*29*, *33*), and/or minor mergers with the accreting satellites (*30*).

76. M. Cappellari, Improving the full spectrum fitting method: accurate convolution with Gauss–Hermite functions. *Mon. Not. R. Astron. Soc.* **466**, 798–811 (2017).

**Acknowledgments:** We thank K. Tadaki, Y. Matsuda, Y. Harikane, D. Iono, I. Mitsuhashi and S. Fujimoto for discussions about high redshift submillimeter galaxies, K. Onishi and D. Nguyen for discussions on the gas dynamics in galaxies, and H. Mikami, J. Ueda and H. Nagai for support of the ALMA data reduction. ALMA is a partnership of ESO (representing its member states), NSF (USA) and NINS (Japan), together with NRC (Canada), MOST and ASIAA (Taiwan), and KASI (South Korea), in cooperation with the Republic of Chile. The Joint ALMA Observatory is operated by ESO, AUI/NRAO and NAOJ. Data analysis was in part carried out on the common-use data analysis computer system at the East Asian ALMA Regional Center (EA ARC) and the Astronomy Data Center (ADC) of the National Astronomical Observatory of Japan (NAOJ). **Funding:** T.T was partially supported by Overseas Travel Fund for Students (2019) of the Department of Astronomical Science, the Graduate University for Advanced Studies, SOKENDAI. **Author contributions:** T.T. led the data analysis. T.T. and S.I. discussed the interpretation of the results and wrote the manuscript. **Competing interests:** We declare no competing interests. **Data and materials availability:** The ALMA pipeline calibrated data and raw data are available at http://almascience.nrao.edu/aq/ under project code ADS/JAO.ALMA #2017.1.00394.S. Part of the ALMA data were retrieved from the JVO portal (http://jvo.nao.ac.jp/portal/) operated by ADC/NAOJ. Supplementary data files contain the observed [CII] data cube cutout used for dynamical modeling (data S1), model [CII] data cube from KINMS (data S2), observed full [CII] data cube used for displaying a spectrum (data S3), and observed dust continuum image (data S4). Our derived model values are listed in Tables S1 and S2.14

**Supplementary Materials:**

Materials and Methods

Figures S1-S6

Tables S1-S2

References (*34-76*)

Data S1 to S4

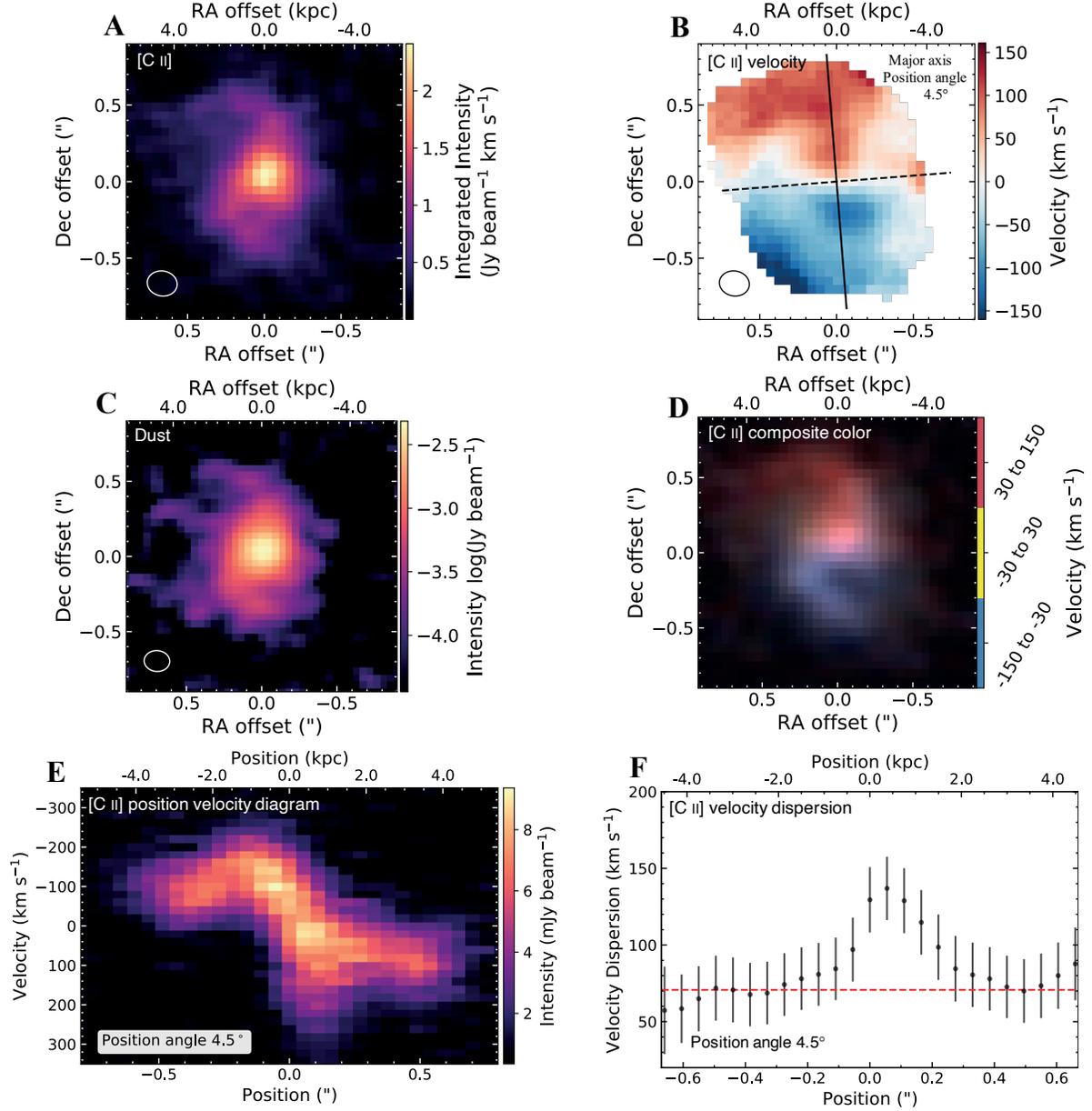

**Fig. 1. Morphology and kinematics of [C II] line and dust continuum emission in BRI 1335-0417.** (**A**) and (**B**) the ALMA [C II] line maps of intensity (moment 0) and intensity-weighted velocity (moment 1). The kinematic major and minor axes are shown in panel B by black solid and dotted lines, respectively. (**C**) the ALMA dust continuum map at a rest-frame frequency of 1863 GHz. (**D**) a composite color image showing in red the redshifted gas emission with velocity of 30 to 150 km s$^{-1}$, in yellow gas at the systemic velocity of -30 to 30 km s$^{-1}$, and in blue the blueshifted gas with velocity of -150 to -30 km s$^{-1}$. The FWHM of the synthesized beam is 0.20" × 0.16" in (A), (B) and (D), and 0.17" × 0.14" in (C), which are shown by an ellipse in the bottom left corners. The RA and Dec offsets are given relative to the position (J2000) 13$^h$38$^m$03$^s$.416, -4°32'35".02 throughout this paper. (**E**) PVD of the observed [C II] line emission, extracted along the kinematic major axis shown in panel B, which has a position angle of 4.5° (Table S1). Data were averaging over one beam width (three pixels). (**F**) the velocity dispersion measured from the line profile at each position of PVD. The error bars correspond to the sum in quadrature of the profile fitting uncertainty (95% confidence level) and the velocity channel width. The red dashed line shows the estimated intrinsic velocity dispersion of $71^{+14}_{-11}$ km s$^{-1}$. Positive position in panels (E) and (F) is the direction toward the north along the kinematic major axis in (B).

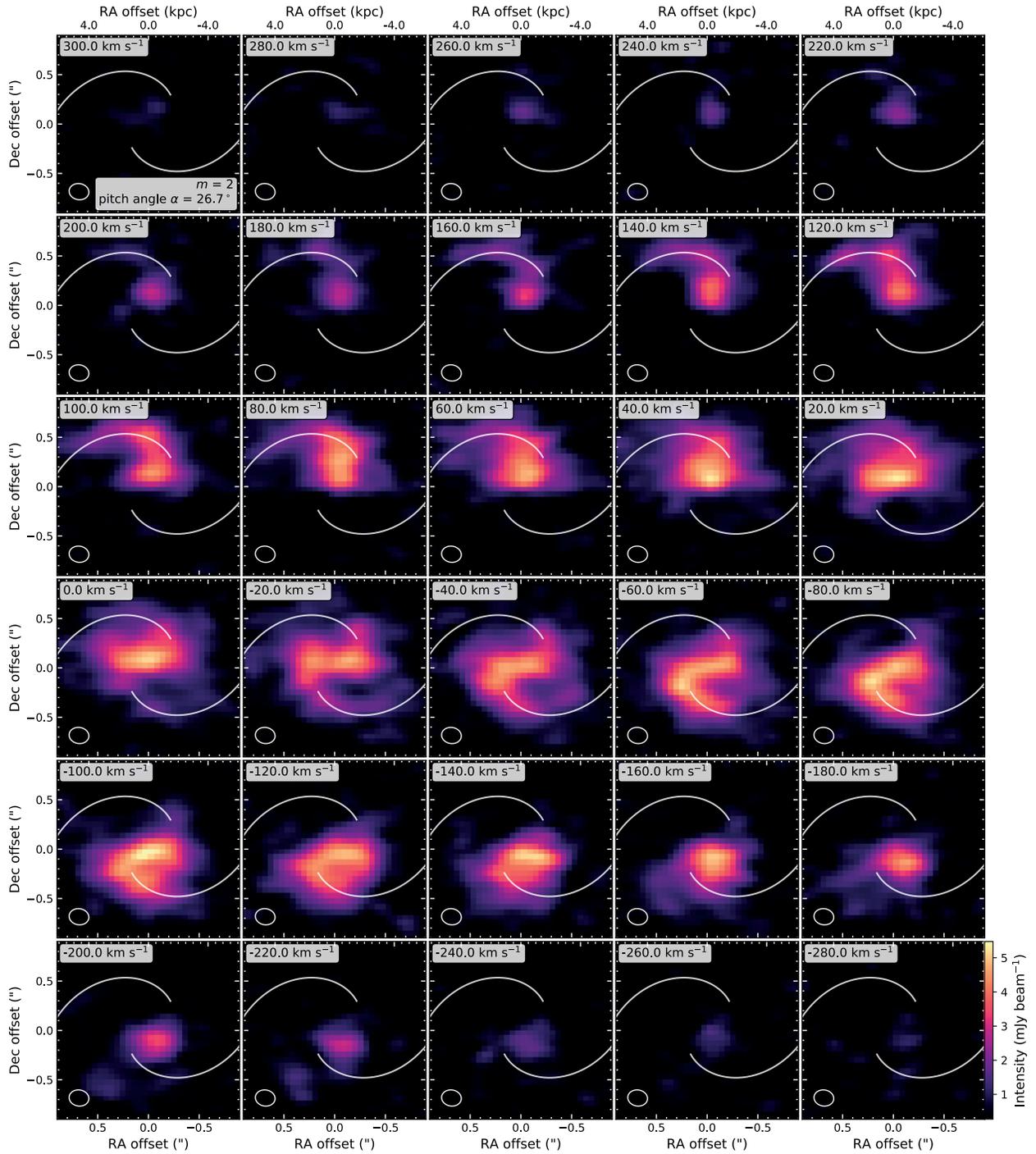

**Fig. 2. Velocity channel maps of the [C II] line emission, showing the spiral structure.** Each panel shows an image at each velocity of the [C II] line emission, labelled in the top left. Overline with white lines is the best-fitting two-armed logarithmic spiral model, which has a pitch angle $\alpha$

= 26.7° (*14*). The images have been de-projected to be viewed as face-on using a position angle of 4.5° and inclination of 37.8° (Table S1) (*14*). The FWHM of the synthesized beam is shown by the white ellipse in the left bottom of each panel.

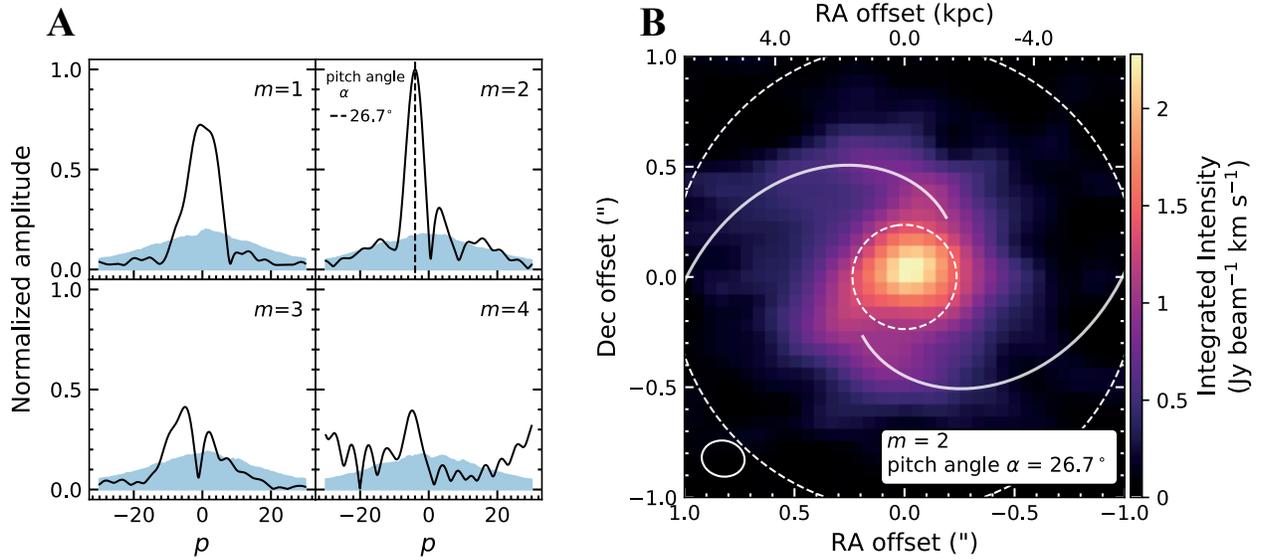

**Fig. 3. Fourier analysis of the [C II] line emission.** (**A**) Fourier spectra (*14*) of model logarithmic spirals with *m* armes and pitch angles *α*. The black solid lines show normalized amplitude as a function of the dimensionless parameter *p* = -*m*/tan(*α*). The blue shaded region indicates the estimated noise level (*14*). The *m* = 2 mode has the strongest peak at a pitch angle of $26.7^{+4.1}_{-1.6}°$ (dashed vertical line). (**B**) the [C II] line intensity image (de-projected version of Fig. 1A) overlaid with the best-fitting two-armed logarithmic spiral (white line, same as in Fig. 2). The FWHM of the synthesized beam is shown by an ellipse in the left bottom corner. The inner and outer dotted lines indicate the boundaries of the region used in the Fourier analysis (*14*).

Supplementary Materials for

Spiral morphology in an intensely star-forming disk galaxy more than 12 billion years ago

Takafumi Tsukui*, Satoru Iguchi

Correspondence to: t.tsukui@grad.nao.ac.jp

**This PDF file includes:**

    Materials and Methods
    Figs. S1 to S6
    Tables S1 to S2
    Captions for Data S1 to S4

**Other Suplementary Materials for this manuscript includes the following:**

    Data S1 to S4

**Materials and Methods**

Cosmology

The conversion from the angular size (arcsec) to the physical size (kpc) is based on a concordance, flat lambda cold dark matter (ΛCDM) cosmology with density parameter of pressureless matter, $\Omega_m = 0.286$ and present-day Hubble constant, $H_0 = 69.6$ km s$^{-1}$ Mpc$^{-1}$ (34), providing the spatial scale of 6.808 kpc arcsec$^{-1}$.

ALMA observation

The rest frame frequency of the [C II] line is 1900.5369 GHz. Because the highest observing frequency of the ALMA is 950 GHz, the [C II] line emission is not detectable from local galaxies. In the distant galaxies ($z > 1$), however, [C II] line is redshifted to the frequency range which can be observed with ALMA. At the redshift for BRI 1335-0417 of $z = 4.4074$ (11), the [C II] line of 1900.5369 GHz is shifted to 351.470 GHz and can be observed with ALMA Band 7 (275 – 373 GHz) that has two single sideband (2SB) receiver systems (35), providing upper sideband (USB) and lower sideband (LSB) simultaneously with the intermediate frequency (IF) bandwidth of 4-8 GHz. We retrieved archival ALMA observations obtained on 2018 January 21 in the observing frequency range of 337.434 GHz -341.433 GHz (LSB) and 349.705 GHz -353.271 GHz (USB), so the USB contains the redshifted [C II] line emission. The LSB consists of two spectral windows, each of which has 128 frequency channels with the widths of 15.625 MHz, while USB consists of two spectral windows, each of which has 1920 frequency channels with the widths of 977 kHz. The total bandwidths of LSB and USB were 4 GHz and 3.75 GHz respectively. The observation was executed in the C43-5 array configuration with baseline lengths of 15 m to 1398 m. The 5th percentile shortest baseline of the configuration was 70 m, providing the maximum recoverable scale of 2.47 arcsecond at 351.47 GHz [(36) their equation 7.7]. Total observing time was 2.0 h including on-source time of 1.0 h and the other 1.0 h for calibration sources and overhead. QSO B1334-127 (VCS1 J1337-1257) was utilized as a flux and bandpass calibrator and NGC 5232 (VCS1 J1336-0829) as a phase calibrator.

Data analysis and imaging

We performed data reduction using version 5.1.1 of the Common Astronomy Software Applications (CASA) (37) pipeline. We first imaged the dust continuum emission shown in Fig. 1C using the line-free channels (with a 4.75 GHz total bandwidth over both side bands at a central frequency of 345 GHz in the observed frame corresponding to 1863 GHz in the rest frame) using the task tclean in CASA. We identified the line-free channels using the hif_findcont task in CASA and additionally removed channels which show a dip feature around ~ 352.3 GHz due to atmospheric absorption. The visibility data were weighted by the Briggs scheme with a robust parameter of 0.0 which provides a trade-off between resolution and sensitivity (38), resulting in a synthesized beam with FWHM 0.17" × 0.14" (~1.16 kpc × 0.95 kpc in physical scale) at a position angle of 86° and a root mean square noise of 0.036 mJy beam$^{-1}$. Secondly, we imaged the [C II] line emission shown in Fig. 1 (A and B) and Fig. S1 (A and D). We measured the flux density of the dust continuum emission by fitting a linear function to the line-free channels and subtracted its emission components from the visibility data using the task uvcontsub in CASA. After continuum subtraction, we imaged two spectral windows of the USB to produce a data cube with the three dimensions of right ascension, declination and velocity using the task tclean in CASA. The velocity channel resolution was 20 km s$^{-1}$ (=23.4 MHz) after binning by 24 frequency channels of 977 kHz, to improve the signal to noise ratio. The visibility data were weighted by the Briggs scheme with a robust parameter of 0.5, resulting in a synthesized beam with FWHM 0.20" × 0.16" (~1.33 kpc × 1.11 kpc in physical scale) at a position angle of 80°. The synthesized beam (i.e. point spread function) is shown in Fig S2; there are negligible side lobes, due to the broad uv coverage of this data. The root mean square noise of each velocity channel with a width of 20 km s$^{-1}$ is about 0.35 mJy beam$^{-1}$. The clean threshold was set to 1.5 times the rms noise for the dust continuum image, and 1.0 times rms noise for [C II] line image, to maximize the flux in the cleaned model. The clean mask was drawn closely around the areas of source emission. For both dust continuum and [C II] line images, the cellsize was chosen to be roughly one third of the minor synthesized beam width. Primary beam correction was performed, which was smaller than 1% at the edge of the images. Figures 1 (A and B) and Figures S1 (A and D) show images of the integrated flux and line-of-sight velocity respectively. Figures 1B and Figures S1D were made by the masked moment method (39). The observed data cube was smoothed by convolving with a uniform filter spatially (1.5 times the beam size) and spectrally (4 times the channel width), then the mask area was determined by the intensity threshold (7.5 times the root mean square noise in the smoothed cube) and applied to the original observed cube. Finally, these velocity images (Moment 1) were created from the masked cube using plotting code in KINMS package (40). The filter size and threshold of the mask were selected so the mask covered the [C II] line emission area.

The size and inclination of the disk
Both the dust continuum and [C II] line intensity distributions of BRI 1335-0417 include complex structures which may indicate a disk and spiral arms. The position angle of the disk was estimated by using PAFIT Package (*41*) from the [C II] line velocity image (Moment 1), which shows the regular rotation and kinematic position angle. The central 0.5" region in [C II] line velocity image was utilized for measuring position angle, as this region features sufficient signal to noise ratio. The estimated position angle is $4.5 \pm 3.7°$ (95% confidence interval). We also derived the position angle of the disk from the gas dynamical modeling (see below), finding $7.6^{+7.3}_{-7.1}$ ° at 95% confidence level. To estimate the axis ratio of the disk with the estimated position angle, we fitted 2-dimensional (2D) Sérsic functions to the dust continuum image (see Fig 1C) and the [C II] line intensity image (Fig. 1A and Fig. S1A), using GALFIT code (*42*). While fitting these images, the synthesized beam was taken into account. From the estimated axis ratio ($q$) of the dust continuum image, the inclination of the disk $i = 37.8^{+2.4}_{-3.3}$ ° was calculated with equation $q = \cos(i)$ and the confidence interval was estimated from the uncertainty of the measured position angle. This estimated inclination was used as a prior in our MCMC sampling (see below), where the gas dynamical modeling finds $37.3^{+3.0}_{-3.1}°$ at 95% confidence level. As GALFIT outputs, the measured Sérsic index is $n = 0.87 \pm 0.02$ (close to exponential $n=1$) from the [C II] line intensity distribution, while it is $n = 2.48 \pm 0.05$ from the distribution of the dust continuum emission. These indicate that the [C II] line distribution has shallow and near exponential profile like a typical disk (*23*, *43*), while the dust continuum distribution has centrally concentrated profile with high Sérsic index ($n > 2$) compatible with that of a typical galaxy with a bulge (*44*). We therefore fitted a 2D exponential function to the [C II] line intensity image, resulting in a disk-scale radius $R_d = 1.83 \pm 0.04$ kpc. The physical parameters derived in this section are listed in Table S1.

Spiral structure
To quantitatively measure the degree of the spiral structure, we decomposed the [C II] line intensity distribution of BRI 1335-0417 into *m*-armed logarithmic spirals, which is expressed as

$$\rho = \rho_0 \exp\left(-\frac{m}{p}(\varphi+\varphi_0)\right) \quad \text{(S1)}$$

in polar coordinates ($\rho$, $\varphi$). $p$ is a variable related to the pitch angle $\alpha$ by $p = -m/\tan(\alpha)$, and $\rho_0$ and $\varphi_0$ are arbitrary constants. We first deprojected the [C II] line intensity image to be viewed as face-on using the position angle 4.5° and inclination 37.8° measured above (Table S1). The image rotation and stretching for the deprojection were performed using SCIKIT-IMAGE package (*45*), in which mapping the original pixels onto a new pixel grid was done by Bi-linear spline interpolation. From the deprojected image, we then calculated the Fourier amplitude $A$ (*3*, *46*), which indicates the magnitude of the *m*-armed logarithmic spirals. It follows that

$$A(p,m) = \frac{1}{D}\sum_{j=1}^{N} f_j \exp\left(-i\left(pu_j + m\varphi_j\right)\right) \quad \text{(S2)}$$

where $D = \sum_{j=0}^{N} f_j$ is the sum of the flux $f_j$ of the pixel $j$ in the image, $u_j$ is the logarithmic radius from the center of the image and $\varphi_j$ is the polar angle. We excluded the central region of 1.6 kpc and the region outside of 7.2 kpc (see Fig. 3B) for the calculation of $A(p,m)$ to avoid the effects of unresolved components and noise. The center of the polar coordinate was defined as the pixel which has a peak intensity in the [C II] line image.

Figure 3A shows the Fourier amplitude $A(p,m)$ as a function of $p$ for $m = 1, 2, 3, 4$. Its underlying noise level was estimated by applying equation (S2) on 300 noise maps extracted from the emission line region of the data cube and computing their 84th percentile. The strongest peak is seen in $m = 2$ at $p = -4.0^{+0.6}_{-0.3}$, corresponding to spiral arms with a pitch angle of $\alpha = 26.7^{+4.1}_{-1.6}°$ at the 95% confidence level. The confidence interval was estimated from the uncertainties of the position angle and inclination of the disk. Figure 3B shows the deprojected [C II] line intensity image by using the derived inclination and position angle. The two-armed logarithmic spirals ($m = 2$) is plotted using equation (S1) with $\rho_0 = 2.2$ kpc and $\varphi_0 = 55°$ in Fig 2 and Fig. 3B (a white solid line).

Gas dynamical modeling
We used the KINMS code (*40*) to produce a gas dynamical model cube which has the same beam size and velocity resolution as those in the observed cube (i.e. beam smearing effect was taken into account for the model cube). It requires a set of gas particles which approximates the distribution of gas intensity in physical space (right ascension, declination and distance along the line of sight), and then the line-of-sight velocity of each particle is calculated by taking a circular velocity of the galaxy as an input. The SKYSAMPLER code (*47*) was used to generate gas particles that reproduce the [C II] line intensity profile of BRI 1335-0417 by using the de-convolved CLEAN components that were provided by a task of tclean in CASA. Then, it transformed the distribution of gas particles from the sky plane into the physical space using the inclination and position angle of the galactic disk.

In general, the structure of disk galaxies consists of a central black hole, a centrally concentrated bulge, an extended disk and a dark matter halo (*48–50*). The circular velocity $v_{circ}(r)$ of the galaxy can be expressed by

$$v_{circ}(r)^2 = v_{BH}(r)^2 + v_{bulge}(r)^2 + v_{disk}(r)^2 + v_{DM}(r)^2 \quad (S3),$$

where $r$ denotes the radial distance from the galaxy center, and $v_{BH}(r)$, $v_{bulge}(r)$, $v_{disk}(r)$ and $v_{DM}(r)$ are the circular velocity of a test particle at $r$ due to the gravitational potential of the central black hole, bulge, disk, and dark matter, respectively. The sphere of influence of the black hole $R_{SOI} = GM_{BH}/\sigma_\star^2 = 0.02"$ where the $G$ is the gravitational constant, $M_{BH}$ is the mass of the black hole and $\sigma_\star$ is the stellar velocity dispersion, was calculated for a mass of $6 \times 10^9\,M_\odot$, which was estimated using the virial theorem in a previous study (*26*), assuming the local relationship between black hole mass and stellar velocity dispersion (*25*). Because estimated $R_{SOI}$ is much smaller than our spatial resolution, we cannot measure the circular velocity of black hole $v_{BH}(r)$ and bulge $v_{bulge}(r)$ separately. We thus included the black hole mass in the bulge. In this case, $v_{bulge}(r)$ is equal to $v_{BH}(r)$ in the limit that the size of the bulge goes to 0, allowing a possibility of a black hole without a bulge. The mass contribution of dark matter is typically smaller than that of the baryonic structure such as the bulge and disk within the radius of < 10 kpc, which are suggested by the recent observation for high redshift galaxy (*50–52*). In this object, the [C II] line emission is detected only up to 5 kpc, which is not sufficient to measure the mass distribution of the dark matter using the kinematics. We therefore excluded the dark matter mass distribution from our dynamical model. A circular velocity $v_{circ}(r)$ of the galaxy is then approximated by

$$v_{circ}(r)^2 = v_{bulge}(r)^2 + v_{disk}(r)^2 \quad (S4).$$

Bulge model
In the PVD shown in Fig. 1E, there is a steep rise of [C II] line velocity reaching about 240 to 260 km s$^{-1}$ in the galactic center, indicating a centrally concentrated mass distribution such as a bulge. Observations of nearby galaxies (*23*, *43*) and distant galaxies ($2 < z < 3$) (*53*, *54*) indicate that the observed surface brightness distributions of a bulge can be approximated by de Vaucouleurs profile (*55*). Assuming the bulge has the same mass profile as its brightness profile, we can express the surface mass density distribution using the de Vaucouleurs profile:

$$\Sigma_{bulge}(R) = \Sigma_e \exp\{-\kappa[(\tfrac{R}{R_e})^{\tfrac{1}{4}} - 1]\} \quad (S5),$$

where $\kappa \sim 7.6695$ (*56*), $R$ is the projected radius on the plane of the sky and $\Sigma_e$ is the surface mass density at the effective radius $R_e$ in which the enclosed mass becomes half of the total mass of the bulge $M_{bulge}$. The total mass of the bulge is then given by $M_{bulge} \sim 22.665 R_e^2 \Sigma_e$ (*49*). In the case of a spherical bulge, the circular velocity of the bulge $v_{bulge}(r)$ at $r$ can be determined by two parameters: the total mass of the bulge $M_{bulge}$ and the effective radius $R_e$. It follows that (*57*, *58*)

$$v_{bulge}^2(r) = -\frac{4G}{r}\int_{l=0}^{r}\left[\int_{R=l}^{\infty}\frac{d\Sigma_{bulge}(R)}{dR}\frac{dR}{\sqrt{R^2-l^2}}\,l^2\right]dl \quad (S6).$$

The circular velocity of spherical structure differs from that of non-spherical one. In the case of a non-spherical structure with an axial ratio greater than 0.6, however, the difference between them is less than 8 percent, allowing us to use spherical assumption for simplicity.

Disk model
An exponential disk profile is predicted by the theory of galactic disk formation (*59*) and observed in nearby galaxies (*23*, *43*) and distant galaxies ($z \sim 2$-3) (*52–54*, *60*). In addition, the exponential profile has been widely used for disk mass distribution in modeling the galactic dynamics including the Milky Way (*49*, *50*, *61*). A surface mass density distribution of an exponential disk is given by

$$\Sigma_{disk}(R) = \Sigma_0 \exp(-R/R_d) \quad (S7),$$

where $R_d$ is a disk-scale radius and $\Sigma_0$ is the central surface mass density. The circular velocity of the disk can be determined from two parameters: the disk-scale radius $R_d$ and the total mass of the disk $M_{disk} = 2\pi\Sigma_0 R_d^2$. It follows that (*62*)

$$v_{disk}(r)^2 = 4\pi G \Sigma_0 R_d\, y^2 [I_0(y)K_0(y) - I_1(y)K_1(y)] \quad (S8),$$

where $y = r/(2R_d)$. $I_0$, $K_0$, $I_1$ and $K_1$ are modified Bessel functions. Several studies suggest that the [C II] line is a good tracer of the total gas mass (*63, 64*). The [C II] intensity distribution of this object is well matched by the exponential profile. We therefore used $R_d = 1.83$ kpc measured by the GALFIT code (Table S1).

Rotation velocity of pressurized galactic disk
The PVD in Fig. 1E shows that BRI 1335-0417 has velocity dispersions of ~70 km s$^{-1}$ in the outer parts. The intrinsic velocity dispersion averaged for galaxies at $z \sim 2.3$ is 50 km s$^{-1}$ but there are a wide range of values up to ~ 80 km s$^{-1}$, and for galaxies at $z \sim 0.9$ is 25 km s$^{-1}$ (*65*). The pressure in the gas disk is proportional to the square of the velocity dispersions of the gas, which is not negligible for the gas dynamical modeling (*61*). The pressure gradient causes a radial force that balances part of the gravitational force, resulting in the reduction of the rotational velocity of the gas. Under hydrostatic equilibrium, the rotational velocity $v_{rot}(r)$ for the exponential gas distribution is given by (*20, 61*)

$$v_{rot}(r)^2 = v_{circ}(r)^2 + 2\sigma_v^2 \times \frac{d\ln\Sigma_{gas}}{d\ln r}(r)$$
$$= v_{circ}(r)^2 - 2\sigma_v^2 \times (\frac{r}{R_d}) \quad \text{(S9)},$$

where $\sigma_v$ is the intrinsic velocity dispersion of the gas, and $\Sigma_{gas}(r)$ is the surface density distribution of the gas disk, which was assumed to be exponential with the disk-scale radius $R_d$ (Table S1). Because the velocity dispersion is constant over this galaxy except in the central region [Fig. 1 (E and F) and Fig. S1(G and H)], we adopted an isotropic and spatially uniform velocity dispersion. This is consistent with previous works (*20, 50, 66*). Given the high velocity dispersion, we considered the disk thickness of $h \sim (\sigma_v/v_{rot}(R_d)) \times 1.68 R_d$ (*50*), correcting equation (S8) (*58*).

MCMC fitting
We determined best-fitting model parameters and their associated confidence intervals with a Markov Chain Monte Carlo (MCMC) sampling using the EMCEE code (*67*), that is an implementation of affine invariant MCMC ensemble sampler. We directly compared the observed cube with the model cube, then derived the following eight physical parameters: the position angle and inclination of disk ($\Gamma$, $i$), the masses of the central compact structure and the disk ($M_{bulge}$, $M_{disk}$), the effective radius of the compact structure ($R_e$) and the offsets between the observed and model cubes in right ascension, declination and velocity ($\Delta x$, $\Delta y$, $\Delta v$). Following the Bayesian framework, the posterior probability distribution of a set of these physical parameters $\theta$ on the data is

$$P(\theta|\text{data}) \propto P(\theta) \times P(\text{data}|\theta) \quad \text{(S10)},$$

where $P(\theta)$ is the assumed prior probability distribution of the physical parameters, and $P(\text{data}|\theta)$ is the likelihood that is the probability of obtaining the data on the given parameters $\theta$. Because the inclination of the disk $i$ was estimated with GALFIT above (Table S1), the prior distribution was set to be a Gaussian, characterized by the estimated central value and the associated uncertainty. We used log uniform prior distribution for the mass of the bulge $M_{bulge}$ and disk $M_{disk}$ to cover the multiple orders of magnitude, while uniform prior distribution for the other 5 parameters. The range was set to values much larger than expected from the data. The likelihood can be written as

$$P(\text{data}|\theta) \propto \exp(-\chi(\theta)^2/2) \quad \text{(S11)}$$
$$\chi(\theta)^2 = \Sigma_{i=0}^{N}[(\frac{I_{\text{data},i} - I_{\text{model},i}(\theta)}{\sigma_{rms}})^2] \quad \text{(S12)},$$

where $I_{\text{data},i}$ and $I_{\text{model},i}(\theta)$ are the intensity of the pixel $i$ in the observed cube and model cube respectively. $\sigma_{rms}$ is the root mean square noise of 0.35 mJy beam$^{-1}$ that was measured using the pixels of only line-free channels in the observed cube under the assumption that it is constant over the cube. The likelihood was computed using the observed cube over the velocity range from -390 km s$^{-1}$ to +370 km s$^{-1}$ with the velocity center 0 km s$^{-1}$ matched to the frequency of 351.470 GHz in barycentric frame.

In the standard $\chi^2$ statistic, the 68% confidence interval corresponds to $\Delta\chi^2 = \chi^2 - \chi_{min}^2 = 1$. A variance of this distribution is $2(N - O)$ where $N$ is the number of constraints and $O$ is the number of inferred parameters. In our data ($N \sim 4 \times 10^4$ and $O = 8$), it becomes $2N$ and yields an unrealistically small confidence interval due to the large number of constraints in fitting the observed cube. We used a method of scaling the standard $\Delta\chi^2$ by a factor of $\sqrt{(2N)}$ (*68*). The scaled loglikelihood by $1/\sqrt{(2N)}$, or equivalently the scaled root mean square noise $\sigma_{rms}$ by $\sqrt[4]{(2N)}$, was used in the previous works (*47, 69, 70*) with the Bayesian framework, giving the same effect by scaling $\Delta\chi^2$. The modified likelihood can be written as

$$P(\text{data}|\theta) \propto \exp(-\chi(\theta)^2/2) \quad \text{(S13)}$$

$$\chi(\theta)^2 = \Sigma_{i=0}^{N} [(\frac{I_{\text{data}, i} - I_{\text{model}, i}(\theta)}{\sigma_{\text{rms}}})^2 / \sqrt{(2N)}] \tag{S14}$$

For our data, we find that MCMC sampling of the posterior distribution converges using this procedure. The correction by the inverse covariance matrix for the pixel to pixel correlations (*71*) in the observed data cube is not taken into account, because it is negligible compared to the effect by scaling the logarithmic likelihood.

Results of the gas dynamical modeling
We run the MCMC fitting with 180 walkers and 4000 steps. Because the first 600 steps are not converged to the posterior distribution (a burn-in phase) by MCMC sampling, these first steps were excluded from the analysis. The outcomes of MCMC sampling are shown in Fig S3, including the one-dimensional histogram of each physical parameter and the covariance maps of the pairs of physical parameters. The best fitting values (median of the marginalized distribution) and their associated 95% confidence intervals are summarized in Table S2. Our gas dynamical modeling provides agreement between the observed cube and the model cube (compared in Fig. S1, A to F). The model PVD is compared to the observed PVD in Fig. S1G, where both PVDs are extracted along the kinematic major axis (Fig. 1B and Fig. S1, D to F). Figure. S1H shows the comparison between the observed velocity dispersion and model velocity dispersion, measured at each position of PVD. The central increase of the observed velocity dispersion is reproduced by the dynamical model under the assumption of the spatially uniform intrinsic velocity.

In Figure 1A and Figure S1A, we can see a spiral morphology in the [C II] line intensity image, but we assumed that the mass distribution of a disk is axisymmetric. This does not affect our mass measurement of the central compact structure because the amplitude of the spiral component is only about 15% of the peak emission at maximum. The two-armed spiral morphology is extended from the radius of 2 kpc to 5 kpc, far from the center. Therefore, the contributions of these two arms to the gravitational potential are small in the center of the galaxy. The two arms are located symmetrically with respect to the galaxy center, so these contributions cancel each other out.

We also consider the possibility that the radiation from active galactic nuclei (AGN) could influence the [C II] intensity distribution, which could bias our estimation of the disk scale radius and our dynamical modeling result. Because the [C II] profile is matched by an exponential profile which is typical to the disk, the influence of the AGN is small. If AGN radiation influenced the [C II] emission, the [C II] emission would become more concentrated, our size estimate of the disk would become smaller and the true disk scale radius would be larger than our estimate. We investigated the dependence of $M_{\text{bulge}}$ and $M_{\text{disk}}$ on the disk scale radius $R_d$ using several different values of $R_d$. To compare with them, we also derived $M_{\text{bulge}}$ and $M_{\text{disk}}$ by adding the disk scale radius as a free parameter in our MCMC fitting. This analysis confirmed that the derived masses of $M_{\text{bulge}}$ is not sensitive to the disk scale radius (see Fig. S4).

CO ($J$=7→6) outflow
A possible bipolar outflow has been reported in CO ($J$=7→6) observations of this galaxy, reaching the velocity of 500 ~ 600 km s$^{-1}$ (*72*). Figure S5 shows the integrated spectrum of [C II] line emission within a circular aperture with the diameter of 2". We defined noise level in the spectrum (σ) by measuring the standard deviation of the line free channels of the integrated spectrum (-990 km s$^{-1}$ to -490 km s$^{-1}$ and 490 km s$^{-1}$ to 990 km s$^{-1}$), resulting in the 1σ noise level of 0.033 Jy beam$^{-1}$. The integrated spectrum is fitted with a single Gaussian (Fig. S5A). In the residual (Fig. S5B), redshifted 1.5σ to 2.5σ emission can be seen over 3 channels (270 km s$^{-1}$ to 330 km s$^{-1}$). When these channels were imaged, a peak is seen on the image. On the other hand, the other moderate emissions in the residual (1 to 2σ) are consistent with noise when viewed on the image plane. Because the position of the redshifted emissions in the image plane coincides with the redshifted part of the disk rotation, corresponding to the peak velocity of the PVD (Fig. 1E and Fig. S1G at about 0.1" and 270 to 330 km s$^{-1}$), the redshifted emissions are likely to originate from a gas clump in the receding part of the rotating disk in the center. We could not find any evidence of bipolar outflow in [C II] line emission, but this does not exclude the bipolar outflow seen in CO ($J$=7→6) line emission because the [C II] line may not be strongly emitted from the outflowing material. For example, outflow is seen in OH line absorption but not in [C II] line emission in some other high redshift galaxies at $z \sim 4$ (*73*). Therefore, we did not consider the outflow kinematics in the gas dynamical modeling.

Toomre *Q* parameter
The dynamical stability of the disk is quantified using the Toomre *Q* parameter. For a gas disk, the parameter is given by (*74, 75*)

$$Q = \frac{\sigma_v \kappa}{\pi G \Sigma_{\text{gas}}} \tag{S15},$$

where $\kappa$ is the epicyclic frequency, $\Sigma_{\text{gas}}$ is the surface density of the gas, and $\sigma_v$ is intrinsic velocity dispersion of the gas. $\kappa$ is given by $\sqrt{2\left(\frac{v_{\text{rot}}^2}{r^2} + \frac{v_{\text{rot}}}{r}\frac{dv_{\text{rot}}}{dr}\right)}$ (*57*). When $Q < 1$, the disk is unstable against the perturbations because the self-gravity of the gas overwhelms the local pressure forces due to the turbulent motion and the differential rotation (*74, 75*). Because the total gas mass is traced by the [C II] line (*63, 64*), we estimated the surface density of the gas $\Sigma_{\text{gas}}$ by using the total molecular gas mass ($M_{\text{gas}} = 5 \times 10^{10}\ M_\odot$) (*16*) and the [C II] line distribution we observed. The rotation velocity was derived from our gas dynamical modeling. Using these parameters, we derived Toomre $Q$ parameter for each spatial pixel (Fig. S6, A and B). The error bars in Fig. S6B reflect the uncertainty of the model rotation velocity at 95% confidence interval.

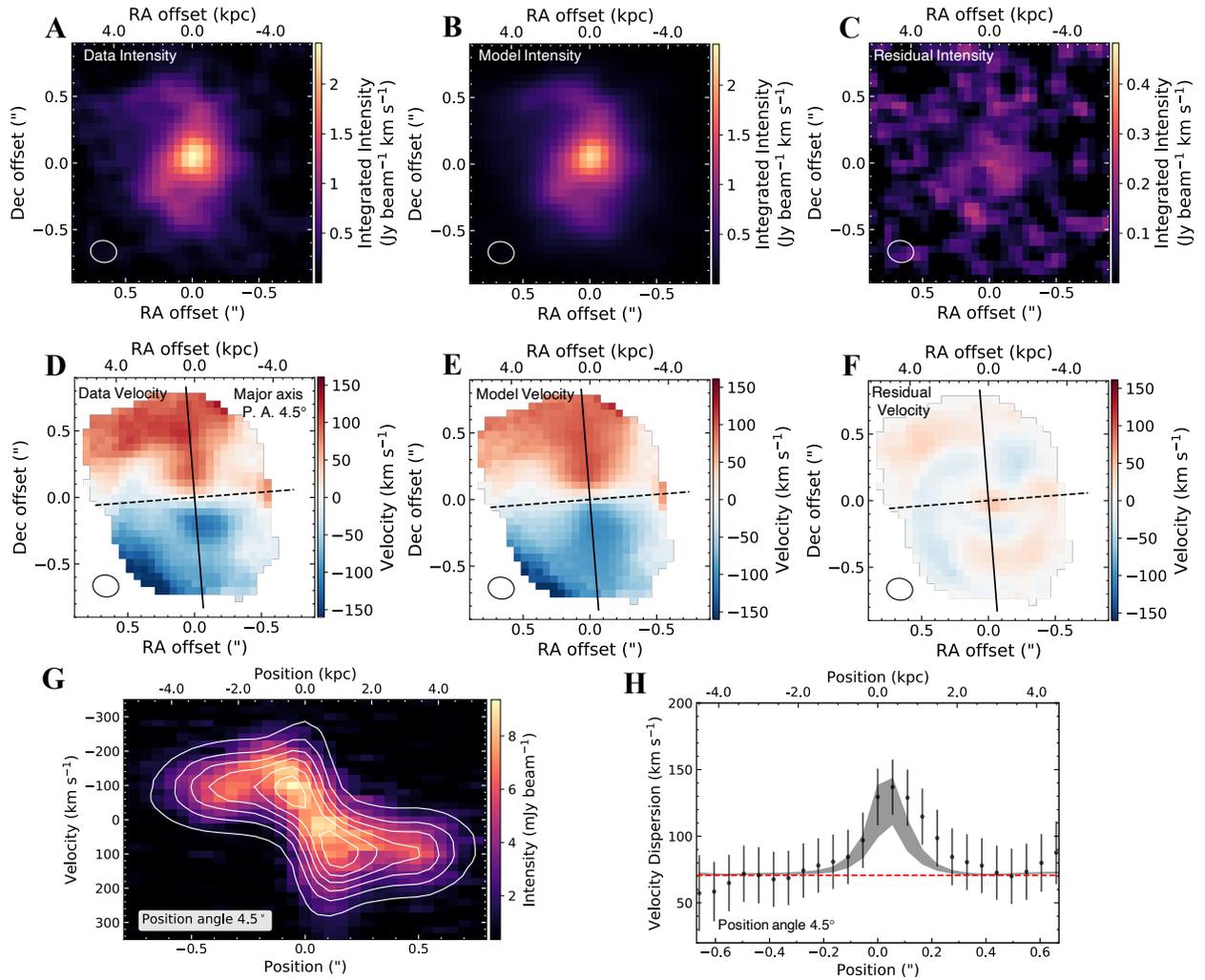

**Fig. S1. Morphology and kinematics of [C II] line in BRI 1335-0417 and their comparisons with gas dynamical models.** (A) and (D) are the respective ALMA [C II] line maps of intensity (moment 0), intensity-weighted velocity (moment 1), same as Figs. 1A and B respectively. (B) and (E) are the dynamical model maps with the central compact mass and disk taken into consideration, which are corresponding to (A) and (D). (C) and (F) are the residual maps between the observed [C II] line maps of (A) and (D) and the dynamical model maps of (B) and (E), respectively. The masked moment method (*39*) was used to produce map (D). The same mask was applied to (E) for a fair comparison. Plotting symbols in (A-F) are the same as Figs. 1A and B. (G) shows the observed PVD of [C II] line emission (color, same as Fig. 1E) and model PVD (white contours: every $4\sigma_{rms}$ from $3\sigma_{rms}$ to $23\sigma_{rms}$, where $\sigma_{rms}$ is the root mean square noise of 0.35 mJy beam$^{-1}$). The model PVD was extracted along the kinematic major axis [see black solid line in (E)]. (H) shows velocity dispersion, which were measured by fitting a Gaussian with third and fourth order Gauss-Hermite parameters $h_3$ and $h_4$ (*76*) to the line profile at each position of PVD (black points and error bars, same as Fig. 1F). This fitting used the pixels where [C II] line emission has signal to noise ratio SNR > 4 in at least 4 velocity channels. (H) also shows velocity dispersion extracted from the model PVD in the same way (gray shade, showing 95% confidence intervals). The red dashed line is same as Fig. 1F.

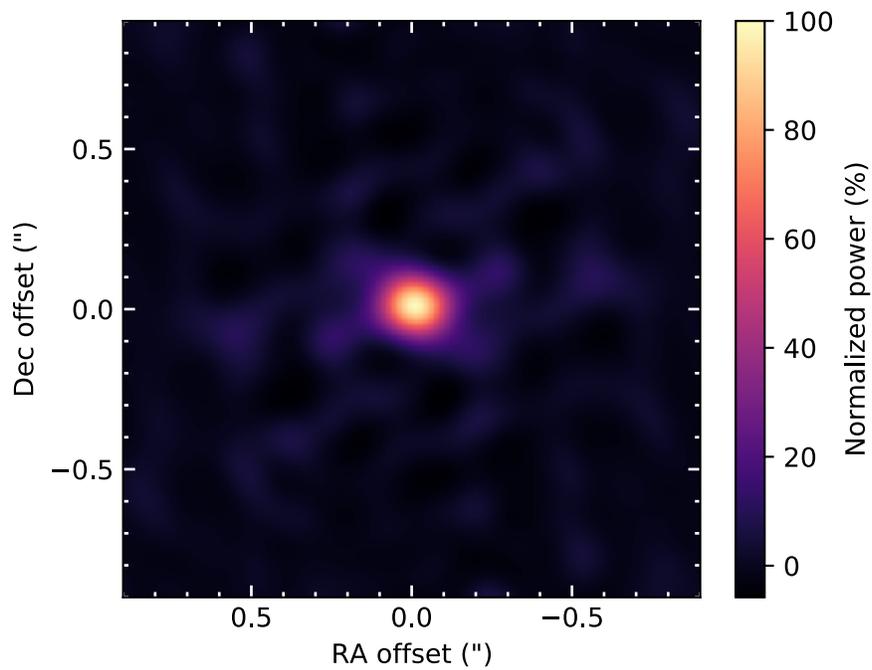

**Fig. S2 Synthesized beam of the ALMA observation.** The synthesized beam (i.e. point spread function) is shown. The side lobes of the synthesized beam are < 13%. The influence of this side lobe was removed from the images by the deconvolution process with a task of tclean in CASA.

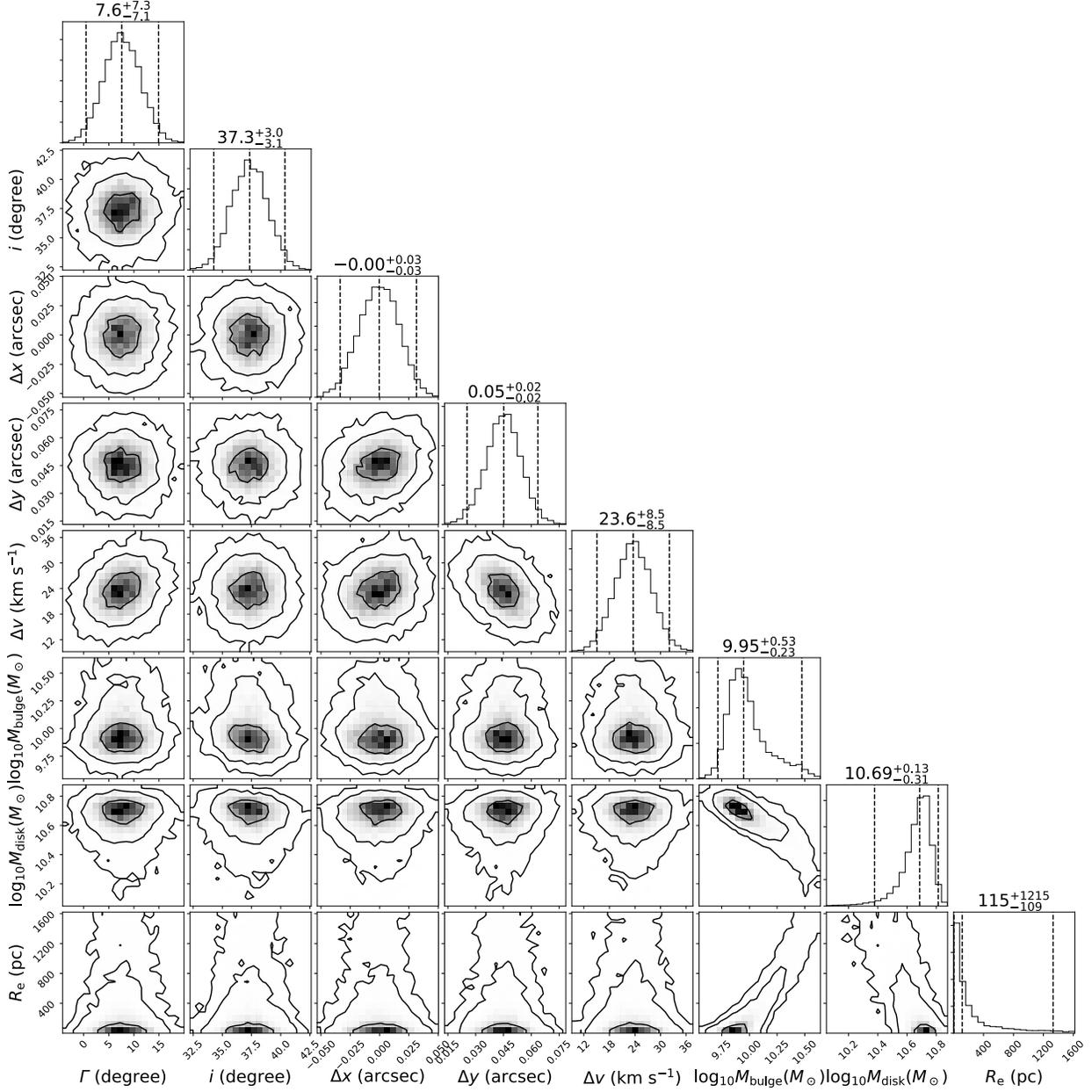

**Fig. S3. Posterior probability distributions of the eight model parameters using MCMC sampling of our gas dynamical model.** The posterior probability distributions for the respective parameters were marginalized, and then the results are shown as the respective histograms along the diagonal panels. The dashed vertical lines in these histograms are the 2.5th, 50th, and 97.5th percentiles (95% confidence interval). The derived values of eight model parameters are indicated above the histogram, summarized in Table S2. The other panels show the covariances between all model parameters (contour levels are equivalent to 68%, 95% and 99.7% confidence intervals for the 2D Gaussian distribution).

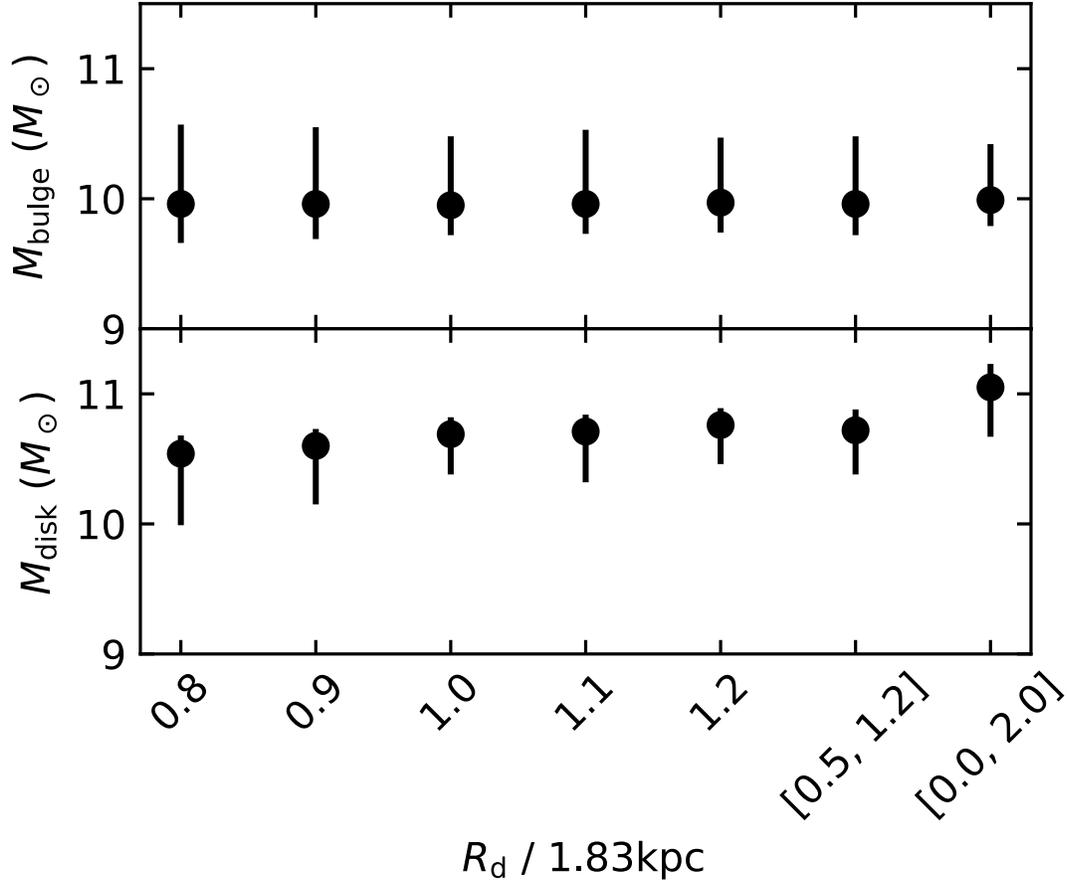

**Fig. S4. The dependency of derived physical parameters on the disk radius.** Best fitting values of $M_{bulge}$ and $M_{disk}$ were derived using several values of the disk scale radius, shown normalized by our estimated value of $R_d = 1.83$ kpc (see Table S1). The disk scale radius was added as a free parameter in the MCMC fitting with the uniform prior range shown in brackets. The derived $M_{bulge}$ is not sensitive to the disk scale radius.

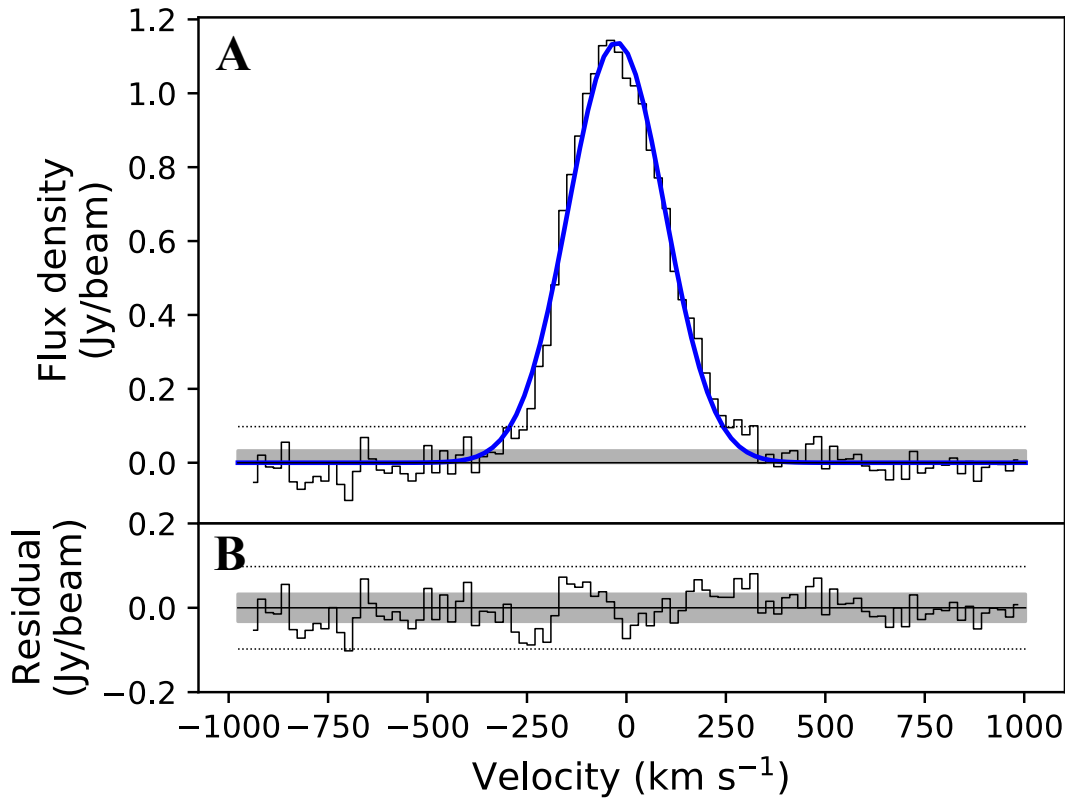

**Fig. S5. Integrated spectrum of the [C II] line emission of BRI 1335-0417.** (A) is the [C II] spectrum (black solid line) extracted using circular aperture with the diameter of 2", which roughly corresponds to the region shown in Fig. S1, A to F. The velocity is relative to the frequency of 351.470 GHz in barycentric frame. A Gaussian model fitted to the spectrum is shown in blue solid line. (B) shows the residual (data – Gaussian model, black solid line). The noise level in the spectrum was estimated by measuring the standard deviation of the line free channels of the integrated spectrum (-990 km s$^{-1}$ to -490 km s$^{-1}$ and 490 km s$^{-1}$ to 990 km s$^{-1}$), resulting in the 1σ noise level of 0.033 Jy beam$^{-1}$. In (A) and (B), the 3σ and 1σ noise levels are indicated by the black dotted line and gray shaded region respectively. The black solid line indicates the zero flux level.

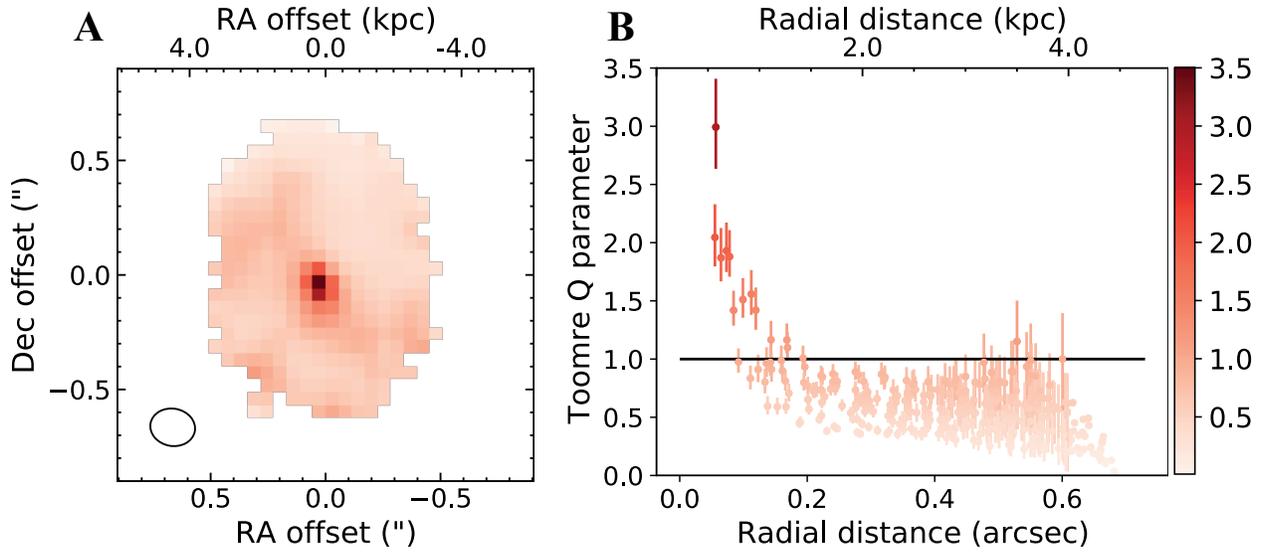

**Fig. S6. The Toomre parameter *Q* for the gas disk of BRI 1335-417.** Spatial distribution of Toomre parameter *Q* for the gas disk is shown in (A). The Q was derived using rotation velocity, intrinsic velocity dispersion and the surface density of the gas. The radial distribution of *Q* is shown in (B), where each pixel in (A) is plotted in radial distance deprojected to a disk plane. Over the outer part of the disk (at the radius of $R > 0.2"$), *Q* is < 1, implying that the gas disk is unstable against the perturbations. The error bars reflect the uncertainty of the 95% confidence interval of the dynamical model. This measurement only used pixels where the [C II] line emission is detected at SNR > 4 in at least 4 velocity channels.

**Table S1.**
**Physical parameter derived by PAFIT Package and GALFIT.** The position angle of the disk was estimated from the [C II] line kinematics using the PAFIT package. The inclination of the disk $i$ was estimated from the axis ratio measured by fitting Sérsic profile to the dust continuum image using the GALFIT code. The reduced chi square $\chi_\nu^2$ of the fitting is 1.592, where the degree of freedom is 4090. Disk scale radius $R_d$ was estimated by fitting exponential profile to the [C II] line image using the GALFIT code. The reduced chi square $\chi_\nu^2$ is 0.956, where the degree of freedom is 4091.

| | |
|---|---|
| Disk position angle $\Gamma$ (°) | $4.5 \pm 3.7$ |
| Disk inclination $i$ (°) | $37.8^{+2.4}_{-3.3}$ |
| Disk scale radius $R_d$ (kpc) | $1.83 \pm 0.04$ |

**Table S2.**
**Best fit summary of our gas dynamical model.** Column 1 shows the parameters and their units in our dynamical model. Column 2 shows the symbol used for each parameter. Column 3 shows the prior distributions used in our MCMC sampling, in which the values in brackets indicate the upper and the lower boundary of the uniform prior distribution, while the values in parentheses indicate the mean and standard deviation of the Gaussian prior. Columns 4 and 5 show the medians of posterior distribution, their associated 95% confidence intervals respectively. The upper limit on the effective radius $R_e$ of the central compact mass, is reported in parentheses. The reduced chi square $\chi_v^2$ in our MCMC fitting is 1.69 under the degree of freedom of 42400.

| Parameters (unit) | Symbols | Priors | Medians | 95% confidence Interval |
|---|---|---|---|---|
| Galactic Mass model: | | | | |
| Central compact mass ($10^9 M_\odot$) | $M_{bulge}$ | [$10^{-9}$, $10^4$] | 8.9 | -3.7, +20.9 |
| Disk mass ($10^{10} M_\odot$) | $M_{disk}$ | [$10^{-10}$, $10^3$] | 4.9 | -2.5, +1.7 |
| Central compact mass effective radius (kiloparsec) | $R_e$ | [0.001, 1.625] | - | (<1.33) |
| Orientation of [C II] gas disk: | | | | |
| Position angle (degree) | $\Gamma$ | [-20, 20] | 7.6 | -7.1, +7.3 |
| Inclination angle (degree) | $i$ | (37.8, ±1.6) | 37.3 | -3.1, +3.0 |
| Nuisance parameters: | | | | |
| [C II] center velocity offset (km s$^{-1}$) | $\Delta v$ | [-60, 60] | 23.6 | ±8.5 |
| [C II] center RA offset (arcsecond) | $\Delta x$ | [-0.15, 0.15] | - 0.00 | ±0.03 |
| [C II] center Dec offset (arcsecond) | $\Delta y$ | [-0.15, 0.15] | 0.05 | ±0.02 |

**Captions for Data**
**S1** Observed [C II] data cube cutout used for dynamical modeling. File in Flexible Image Transport System (FITS) format.
**S2 Model [C II] data cube from KINMS. File in FITS format.**
**S3 Observed full [C II] data cube used for displaying a spectrum. File in FITS format.**
**S4 Observed dust continuum image. File in FITS format.**